\documentstyle[epsfig,aps,pra,twocolumn,floats]{revtex}
\newcommand{\beq}{\begin{equation}}
\newcommand{\enq}{\end{equation}}

\begin{document}


\draft

\wideabs{

\title{Bose-Einstein condensation in shallow traps}

\author{J.-P. Martikainen}

\address{Helsinki Institute of Physics, PL 9, FIN-00014 Helsingin yliopisto,
Finland}

\date{\today}

\maketitle

\begin{abstract}
	In this paper we study the properties of Bose-Einstein condensates 
in shallow traps. We discuss the case of a Gaussian potential,
but many of our results apply also to the traps having a small
quadratic anharmonicity. We show the errors introduced when a Gaussian
potential is approximated with a parabolic potential, these errors
can be quite large for realistic optical trap parameter values. We study the
behavior of the condensate fraction as a function of trap depth and 
temperature and calculate the chemical potential of the condensate in a 
Gaussian trap. Finally we calculate the frequencies
of the collective excitations in shallow spherically symmetric and 1D traps.
\end{abstract}
\pacs{PACS numbers: 03.75.Fi, 32.80.Pj, 03.65.-w}}

\narrowtext

\section{Introduction}
	An atomic Bose-Einstein condensate~\cite{Anderson95,Bradley95,Davis95}
is an excellent tool
for studying quantum many body phenomena, such as collective excitations.
The particle interactions in these condensates are weak and thus 
even quantitave agreement between
theory and experiments can be found. Recently a sodium condensate
was trapped in an optical dipole trap~\cite{Stamper98a}. 
Usually the potential of such trap is approximated by the
parabolic potential, but the shallow nature of the optical
dipole trap makes it clear that for suitably large condensates, the 
anharmonicity of the potential has to be taken into account. 

In this paper we study how the shallowness of the potential changes the 
condensate properties. 
To enable an analytical approach we assume a Gaussian 
potential~\cite{Gaussian_comment}, but we expect that the qualitative features
are valid also for other types of shallow traps.  
Bose-Einstein condensation in power-law potentials has been studied 
previously~\cite{Bagnato87},
but usually the potential has been assumed to be  parabolic or absent.
In Sec. \ref{Cond_frac} we calculate how the condensate 
fraction behaves as a function of temperature 
and how the parabolic approximation may underestimate the BEC transition
temperature considerably.
In Sec. \ref{Chem_pot} we calculate the chemical potential 
and estimate how many condensate particles can be trapped. 
Frequencies of the collective excitations are calculated in Sec. 
\ref{Coll_ex}, and some concluding remarks are given in Sec.
\ref{Conclusions}.

\section{Condensate fraction}\label{Cond_frac}
Particle interactions have
a dramatic effect for particle density distribution, 
chemical potential and collective
excitations of the condensate, but the transition temperature and the 
condensate fraction can be accurately
calculated with the ideal gas model~\cite{Ensher96}. If there is a repulsive
interaction between the trapped atoms, and the chemical potential is almost
the same as the trap depth, we expect changes to the ideal gas results, but
for now, the simple ideal gas model will suffice.
In a harmonic trap the transition temperature $T_c$ is given by
\beq
	k_BT_c=\hbar\left(\frac{N\omega_x\omega_y\omega_z}
	{\zeta(3)}\right)^{1/3},
\enq
where $\omega_\alpha$ is the trap frequency in direction $\alpha$ and
$N$ is the particle number. The condensate fraction is then given 
by~\cite{Giorgini97} 
\beq
	\frac{N_c}{N}=1-\left(\frac{T}{T_c}\right)^3.
\enq 

In a Gaussian potential
\beq
	V(x,y,x)=V_1\left[1-\exp\left(-\frac{x^2}{2\sigma_x^2}-
	\frac{y^2}{2\sigma_y^2}-\frac{z^2}{2\sigma_z^2}\right)\right]
\enq
condensate fraction can behave in a qualitatively different way.
Using a phase space density~\cite{Chempot_comment}
\beq
	f({\bf r},{\bf p})=\frac{1}{\exp(\beta H)-1},
\enq
where $H$ is the total energy, we can calculate the number of 
thermal particles as
\beq
	N_T=\frac{1}{(2\pi\hbar)^3}\int\int d{\bf p}d{\bf r} f(\bf{r},\bf{p}).
\enq
Our trap has a finite depth and therefore we must introduce an appropriate
$r$ dependent cut-off for the kinetic energies. After a short calculation
we see that the number of thermal atoms is given by
\begin{eqnarray}
\label{Nt}
	N_T&=&\frac{16}{\pi}
	\left(\frac{mV_1}{\hbar^2}\right)^{3/2}\sigma_x\sigma_y\sigma_z
	\nonumber \\ 
	&&\int_0^\infty dx\int_0^{s_{max}}ds\frac{x^2s^2}
	{\exp(\beta V_1(s^2+1-\exp(-x^2)))-1},
\end{eqnarray}
where $s_{max}=\exp(-x^2)$.

	When the trap has only a few eigenstates,
the continuum approach fails and we should model the system using a discrete
spectrum. If the potential is spherically symmetric and is approximated as 
a parabola, then the number of different energy levels is roughly 
$V_1/\hbar\omega=\sqrt{V_1 m\sigma^2/\hbar^2}$. This provides a
lower limit to the number of energy levels, since for a Gaussian trap the
separation between adjacent levels becomes smaller as the energy increases
(approaching zero as the energy approaches the trap depth).
If the trap depth is very small, $V_1\approx 9\,\rm{nK}$ 
and $\sigma=15\,\rm{\mu m}$ the trap has roughly ten energy levels.
We expect that for deeper traps the continuum approach should give accurate
results.

\begin{figure}[bht]
\centerline{\epsfig{file=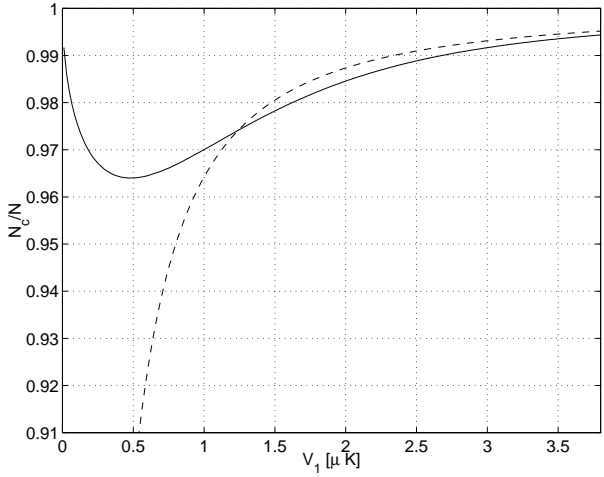,width=8.0cm}}
\vspace*{1cm}
\caption[fig1]{
The condensate fraction as a function of trap depth for
$10^6$ sodium atoms, when $(\sigma_x\sigma_y\sigma_z)^{1/3}=15\,\rm{\mu m}$
and temperature $T=300\,\rm{nK}$. The solid line is the numerically calculated
value using Eq.~(\ref{Nt}) and the dashed line is the result we get
by approximating the Gaussian potential as parabolic.
\label{T300nK}}
\end{figure}

	Let us now consider consider $10^6$ sodium atoms in a trap with 
$(\sigma_x\sigma_y\sigma_z)^{1/3}=15\,\rm{\mu m}$. As the first example
we fix the temperature to the value $T=300\,\rm{nK}$ and vary the
trap depth. In Fig.~\ref{T300nK} we show the resulting condensate fraction
and compare it to the result we get by approximating the potential as
parabolic. At large trap depth the condensate fraction is well predicted
by the parabolic model, but at smaller trap depths the behavior changes
qualitatively. Parabolic model predicts that the condensate fraction
should vanish, together with the trap frequency. This is obviously not
the case. As the trap depth becomes smaller, the condensate fraction takes
a minimum value after which it approaches unity. 
The minimum condensate fraction is achieved when
the trap depth is about the same as the temperature.
This happens because in
a shallow potential there is simply no room for thermal atoms. At
the ultimate limit the potential would have only one bound state and
there would not be any states accessible to thermal atoms.
We also see that at larger trap depths the 
condensate fraction is smaller than the result for the parabolic trap;
a sensible result since anharmonicity makes the trap more open than a purely
parabolic result, thus reducing the effective trap frequency and 
lowering the critical temperature.

\begin{figure}[bht]
\centerline{\epsfig{file=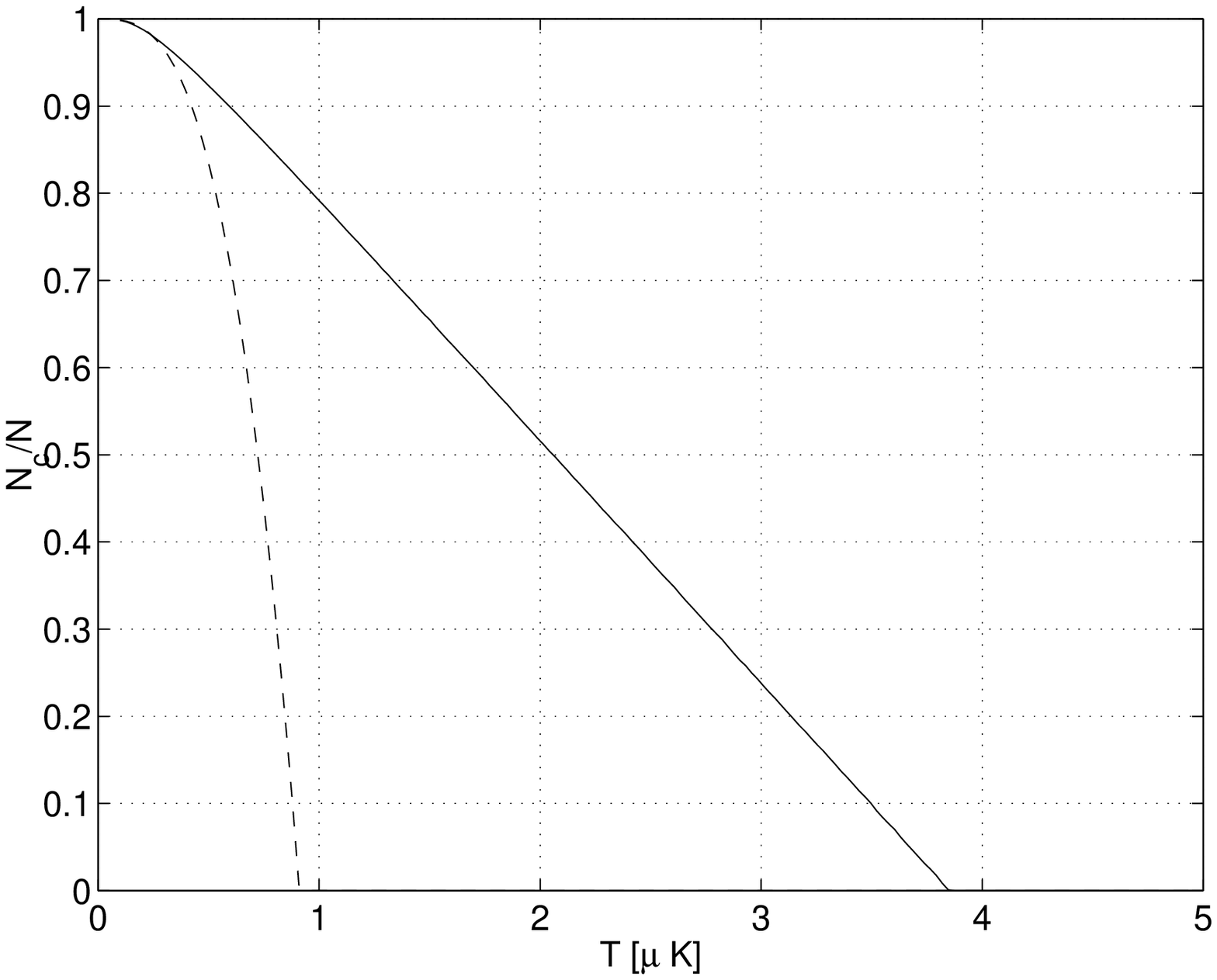,width=8.0cm}}
\vspace*{1cm}
\caption[fig2]{
The condensate fraction as a function of temperature for
$10^6$ sodium atoms, when $(\sigma_x\sigma_y\sigma_z)^{1/3}=15\,\rm{\mu m}$
and trap depth $V_1=1\,\rm{\mu K}$. The solid line is the 
numerically calculated
value using Eq.~(\ref{Nt}) and the dashed line is the result we get
by approximating the Gaussian potential as parabolic.
\label{fraction_T}}
\end{figure}

As a second example we  fix the trap depth $V_1=1\,\rm{\mu K}$ and vary
the temperature. In Fig.~\ref{fraction_T} we show the condensate fraction
and compare it to the parabolic result. Again it can be seen that the
behavior is dramatically different in a shallow potential. The condensate
fraction is considerably larger over a wide range of temperatures and
the critical temperature is about four times larger than the value predicted
by the parabolic model.  
At very small temperatures ($T<240\,\rm{nK}$) the openness of the Gaussian trap
is reflected in condensate fractions, which are smaller than the parabolic
trap predictions. But this effect is so small that it is not visible in
Fig.~\ref{fraction_T}.

\section{Chemical potential versus particle number}\label{Chem_pot}
Properties of pure condensates are well described by the  Gross-Pitaevskii 
(GP) equation
\beq
\label{GP}
-\frac{\hbar^2}{2m}\nabla^2\Psi+V\Psi+NU_0|\Psi|^2\Psi=\mu\Psi.
\enq	
Here $\Psi$ is the condensate wavefunction, $m$ is the atomic mass, $V$ is the
trap potential,
$N$ is the number of atoms and $U_0=4\pi\hbar^2a_s/m$, 
where $a_s$ is the $s$-wave scattering length. 
When the particle number $N$ is large, the kinetic energy becomes small 
compared
to trapping and the atomic interaction energies. 
In the Thomas-Fermi (TF)
approximation the  kinetic energy is ignored and we get an analytical result 
for the condensate wavefunction 
\beq
\label{tf_result}
	|\Psi(x,y,z)|^2=\frac{1}{NU_0}\cdot\left(\mu-V\right),
\enq
when the R.H.S. is positive and zero elsewhere. The wavefunction is 
normalized to 
unity, so by integrating Eq. ~(\ref{tf_result}) we get a formula that 
relates the number of particles to the trap geometry and to the chemical 
potential
\begin{eqnarray}
\label{nmu3d}
	&&\frac{NU_0}{8\pi\sqrt{2}V_1\sigma_x\sigma_y\sigma_z}
	=\frac{x-1}{2}\left[
	\sqrt{-\ln\left(1-x\right)}\,+\right.
	\nonumber\\
	&&\left. \frac{2}{3}\left(-\ln\left(1-x\right)\right)^{3/2}\right]+ 
	\frac{\sqrt{\pi}}{4}\rm{erf}
	\left(\sqrt{-\ln\left(1-x\right)}\right), 
\end{eqnarray}
where error function is defined as
\begin{equation}
	\rm{erf}(t)=\frac{2}{\sqrt{\pi}}\int_0^te^{-s^2}ds
\end{equation}
and $x=\mu/V_1$.
It is clear that the result converges only if $0\le \mu\le V_1$. When 
$\mu=V_1$ we cannot add any more particles to the condensate, as these
extra particles can not be trapped. From this condition we get
the maximum condensate particle number in a Gaussian trap as
\begin{equation}
\label{nctf}
	N_c=\frac{V_1\sigma_x\sigma_y\sigma_z(2\pi)^{3/2}}{U_0}.
\end{equation}
It is instructive to calculate this number for a sodium ($a_s=2.75\, \rm{nm}$ 
\cite{Tiesinga96}) condensate with the reasonable trap parameters 
$V_1=4\,\mu\rm{K}$, $\sigma_x=\sigma_y=3\,\mu\rm{m}$, and
$\sigma_z=38\,\sigma_x$~\cite{Stamper98a}.
The maximum number of condensate atoms is then very large,
\begin{equation}
	N_c\approx 9\cdot10^7.
\end{equation}
For this type of trap the maximum condensate number density would be about
\begin{equation}
	n_{max}=V_1/U_0\approx 5.5\cdot10^{15}\, \rm{cm}^{-3},
\end{equation}
and the three-body decay would limit the condensate lifetime considerably.
It should be noted that for condensates with the maximum number of atoms,
the maximum density depends only on the trap depth and interaction parameter 
and is independent of the trap geometry. This is a general result
and applies to any shallow trap as long as the TF approximation is valid.

	Equation (\ref{nmu3d}) is somewhat awkward and it is useful to
derive a simple approximation for it. The chemical potential is often much 
lower
than the trap depth and we can then expand Eq.~(\ref{nmu3d}) 
around the small parameter
$x=\mu/V_1$. Keeping terms up to order $x^{5/2}$ we get the result 
\beq
	\mu_0=\left(\frac{15NU_0}{16\pi\sqrt{2}V_1\sigma_x\sigma_y\sigma_z}
	\right)^{2/5}\cdot V_1,
\enq
which corresponds to approximating the potential as parabolic and 
reduces
to the familiar formula for parabolic potentials when we notice that the
trap frequencies are related to $\sigma$ and $V_1$ by 
$\omega=\sqrt{V_1/m\sigma^2}$. 

Keeping terms up to order $x^{7/2}$ we get an equation
\beq
\label{x_eq}
	\frac{2}{15}x^{5/2}+\frac{1}{14}x^{7/2}=
\frac{NU_0}{8\pi\sqrt{2}V_1\sigma_x\sigma_y\sigma_z}.
\enq
The solution to this equation is expected to be close to the first 
approximation so we can set $x=x_0+\delta x$ ($x_0=\mu_0/V_1$) and
solve for $\delta x$. The chemical potential is then given by 
\beq
\label{chem_pot}
	\mu=\mu_0\left[1-\frac{3}{14}\left(\frac{\mu_0}{V_1}\right)\right].
\enq 
We have compared this formula to the exact numerical result for a condensate
in a spherically symmetric trap and noticed that
it is an excellent approximation even for relatively large values of $x$.
For example when $x\approx 0.5$ Eq.~(\ref{chem_pot}) is exact with 
a relative accuracy better than $10^{-3}$.

Let us calculate how large this shift in $\mu$ is for some reasonable
parameters. The condition $x_0\ll 1$ implies that
\beq
	\frac{V_1\sigma_x\sigma_y\sigma_z}{N}\gg \frac{15 U_0}{16\sqrt{2}\pi},
\enq
must be satisfied. For a sodium condensate this means
that 
\beq
	\frac{V_1\sigma_x\sigma_y\sigma_z}{N}\gg 2\cdot 10^{-51} \,\rm{Jm}^3,
\enq
a condition that is not difficult to achieve.
Choosing reasonable parameters $\sigma_x=\sigma_y=3\,\mu\rm{m}$, 
$\sigma_z=38\,\sigma_x$, $V_1=4.0\, \rm{\mu K}$ and $N=10^6$ 
\cite{Stamper98a}
we see that Eq.~(\ref{chem_pot}) predicts the shift in the 
chemical potential
to be about $-6\,\%$. This shift is quite large, and it might be necessary to
take this shift into account, when studying condensates in realistic optical
dipole traps.

\section{Collective excitations}\label{Coll_ex}
	In the following subsections we calculate the frequencies of the 
collective excitations in spherically symmetric and 1D traps. We aim
at simple analytical results that clarify the role of trap anharmonicity
and therefore we do not study anisotropic traps. The collective excitation
frequencies for anisotropic traps can be solved numerically,
but analytic results are exceedingly difficult to obtain.
 
\subsection{Spherical trap}
Collective excitations for a spherically symmetric parabolic trap with 
trap frequency $\Omega$
have been calculated in the TF limit analytically~\cite{Stringari96b}.
These excitations have the form $\delta n(r)=P^{2n_r}(r/R)r^l
Y_{lm}(\theta,\phi)$ where $P^{2n_r}$ are polynomials of degree $2n$ and
the dispersion law is given by the formula
\beq
	\omega_0(n_r,l)=\Omega(2n_r^2+2n_rl+3n_r+l)^{1/2}.
\enq
In an anharmonic trap these frequencies will be shifted, but it is not
known by how much. We assume spherically symmetric potential 
\beq
	V(r)=\frac{m\Omega^2}{2}r^2+\Delta V(r),
\enq
where $\Delta V(r)=a r^4$. 
In particular, if the exact trapping potential has the Gaussian shape
\beq
	V_{exact}(r)=V_1\left(1-\exp(-r^2/2\sigma^2)\right),
\enq
we can approximate it with a Taylor series and get 
\beq
	\Omega=\sqrt{\frac{V_1}{m\sigma^2}}
\enq
and
\beq
	\Delta V(r)=\frac{V_1}{8}\left(\frac{r}{\sigma}\right)^4.
\enq

We choose the unit of length to be $L=\sqrt{\hbar/m\Omega}$ and unit
of time as $\tau=1/\Omega$. In these new units the GP-equation becomes
dimensionless
\beq
\label{gp_dimless}
	-\frac{1}{2}\nabla^2\Psi+V(r)\Psi+Na|\Psi|^2\Psi=
	\mu\Psi,
\enq
with the dimensionless interaction parameter $a=4\pi a_s/L$ and potential
\beq
	V(r)=\frac{1}{2}r^2-\epsilon r^4,
\enq
where $\epsilon=\hbar\Omega/8V_1$.

Following Stringari~\cite{Stringari96b} we write the wavefunction in terms 
of phase and modulus,
\beq
	\Psi(\bar{r},t)=\sqrt{n(\bar{r},t)/N}e^{i\phi(\bar{r},t)},
\enq
where $n$ is the density and velocity is fixed by the relation
\beq
	v(\bar{r},t)=(\hbar/m)\nabla\phi.
\enq
The GP equation is equivalent with two coupled equations
\beq
\label{cont_eq}
	\frac{\partial}{\partial t}n+\nabla\cdot(\bar{v}n)=0
\enq
and
\beq
	m\frac{\partial}{\partial t}\bar{v}+\nabla\left(
	V(r)+U_0n-\frac{\hbar^2}{2m\sqrt{n}}\nabla^2\sqrt{n}+\frac{mv^2}{2}
	\right)=0.
\enq
In the Thomas-Fermi limit the kinetic pressure term can be neglected in the
equation for the velocity field, which then becomes
\beq
\label{vel_eq}
	m\frac{\partial}{\partial t}\bar{v}+\nabla\left(
	V(r)+U_0n+\frac{mv^2}{2}\right)=0.
\enq
If we set $\sigma_\alpha=\sigma$ ($\alpha=\rm{x,y,z}$) and the chemical 
potential is sufficiently small, the stationary solution of Eqs. 
(\ref{cont_eq}) and (\ref{vel_eq}) coincide with the Thomas-Fermi 
wavefunction studied in the previous 
section. Linearizing Eqs. (\ref{cont_eq}) and (\ref{vel_eq}) by
setting $n=n_0+\delta n(r)e^{i\omega t}$
gives us an equation for the time-dependent solutions
\beq
\label{omega_eq}
	\nabla\cdot\left[c^2(r)\nabla\delta n\right]+\omega^2\delta n=0.
\enq
Here $n_0$ is the stationary solution and $mc^2(r)=\mu-V(r)$. We now 
calculate the corrections to the
collective excitation frequencies as we move from a parabolic trap into 
an asymmetric trap. We proceed as in the Ref.~\cite{Stringari96b}.

In our case $mc^2=\mu-\frac{m\Omega^2}{2}r^2+\frac{V_1}{8\sigma^4}{r^4}$.
The edge $R$ of the condensate is defined by
\beq
	\mu=\frac{m\Omega^2}{2}R^2+\frac{V_1}{8\sigma^4}{R^4}.
\enq
With this definition the equation (\ref{omega_eq}) takes the form
\begin{eqnarray}
	&\nabla\cdot&\left\{\left[\left(R^2-r^2\right)-
	\frac{V_1}{4\sigma^4\Omega^2m}
	\left(R^4-r^4\right)\right]\nabla\delta n\right\} \\ \nonumber
	&&+2\left(\frac{\omega}{\Omega}\right)^2\delta n=0.
\end{eqnarray}
We choose the dimensionless distance as $x=r/R$  and frequency 
$\omega_D^2=2(\omega/\Omega)^2$. If we postulate a solution 
$\delta n=P(r)Y_{lm}(\theta,\phi)$ and define a dimensionless parameter
$\beta=\frac{R^2}{4\sigma^2}=\mu/2V_1+{\mathcal O}((\mu/V_1)^2)$, 
we obtain an equation for the radial part 
\begin{eqnarray}
\label{P_eq}
	\frac{\partial}{\partial x}&&\left\{x^2\left[(1-x^2)-\beta(1-x^4)
	\right]\frac{\partial P}{\partial x}\right\}\nonumber \\
	&&+\omega_D^2x^2P-l(l+1)\left[(1-x^2)-\beta(1-x^4)\right]P.
\end{eqnarray}
We can consider the terms proportional to $\beta$ as a perturbation and
try to find a first order solution to the equation with form
\beq
	\hat{H}P+\widehat{\Delta H}P=-\omega_D^2P,
\enq
where 
\beq
	\widehat{\Delta H}P=
\frac{\beta}{x^2}\left[l(l+1)(1-x^4)P-\frac{\partial}{
\partial x}\left[x^2(1-x^4)\frac{\partial P}{\partial x}\right]\right]
\enq
and $P$ is the known solution without anharmonicity.
The frequency $\omega$ of the collective excitation is then given by
\beq
	\left(\frac{\omega}{\Omega}\right)^2=(\omega_{0}^2+1/2\cdot\delta
\omega_D^2), 
\enq
where
\beq
	\delta\omega_D^2=\frac{-\int x^2P
	\widehat{\Delta H}P dx}{\int P^2 x^2 dx}
\enq
and $\omega_{0}^2=2n_r^2+2n_rl+3n_r+l$ are the unperturbed frequencies.
As the result is perturbative and expected to be relevant only for the lowest
excitations we will give results only for surface excitations ($n_r=0$) and
the breathing mode ($n_r=1$ and $l=0$).

	For surface excitations ($P=x^l$) we get the result
\beq
\label{surface_ex}
	\delta\omega_D^2=\frac{-4\beta l(2l+3)}{2l+5}
\enq
and for the breathing mode ($P=1-\frac{5}{3}\,x^2$) we get
\beq
\label{Breathing_mode}
	\delta\omega_D^2=\frac{-140\beta}{9}.
\enq

\begin{figure}[bht]
\centerline{\epsfig{file=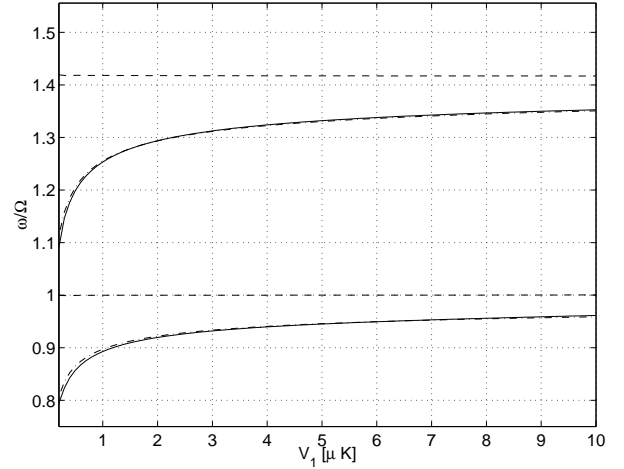,width=8.0cm}}
\vspace*{1cm}
\caption[fig3]{
Frequencies of the collective excitations with 
$(n_r=0, l=1)$ and $(n_r=0, l=2)$ as a function of trap depth,
for $N=10^6$ sodium atoms and 
$\sigma= 15\, \rm{\mu m}$. The solid line is the exact numerically calculated 
value, the dashed line is the numerically calculated result for the
parabolic approximation, and the dot-dashed line is 
based on Eq.~(\ref{surface_ex}). The exact result
and the approximation (\ref{surface_ex}) give almost the same results for
most values of $V_1$.
\label{L12_frequencies_pic}}
\end{figure}

	In Fig.~\ref{L12_frequencies_pic} we compare the exact numerical 
solution,
to the approximation (\ref{surface_ex}), and to the Thomas-Fermi result
for the modes $(n_r=0, l=1)$ and $(n_r=0, l=2)$. We assume $10^6$ sodium
atoms in a trap with $\sigma=15\, \rm{\mu m}$ and vary the trap depth.
We see that our analytical results match the exact numerical values quite
accurately. Since the TF approximation was 
always justified, the numerical values for the parabolic case are 
well predicted with the well known analytical results. If, on the other hand, 
number of
particles would have been less, we would expect $(n_r=0, l=2)$ mode frequency
to decrease significantly with increasing trap depth, approaching the 
value $\sqrt{2}$ asymptotically.
This trend is opposite to the results from a fully Gaussian trap and is a
clear indication of the failure of the parabolic approximation.

\begin{figure}[bht]
\centerline{\epsfig{file=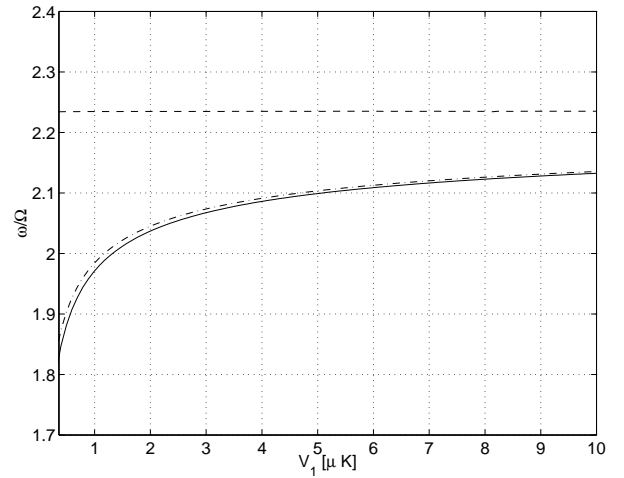,width=8.0cm}}
\vspace*{1cm}
\caption[fig4]{
Frequency of the breathing mode $(n_r=1, l=0)$ as a function of trap depth 
for $N=10^6$ sodium atoms and $\sigma= 15\, \rm{\mu m}$. 
The solid line is the exact numerically calculated value,
the dashed line is the numerically calculated result for the parabolic
approximation, and the dot-dashed line is based on  
Eq.~(\ref{Breathing_mode}).
\label{Breathing_mode_pic}}
\end{figure}

In Fig.~\ref{Breathing_mode_pic} we show the corresponding comparison
for the lowest compressional mode, the breathing mode. Again it can be
seen that Eq.~(\ref{Breathing_mode}) gives an accurate approximation to the
exact result and that corrections to the parabolic model are noticeable.

\subsection{1D trap}
	Spherical symmetry, discussed in the previous subsection, is a special
case not necessarily easily obtained in experiments. To enable analytical 
approach and to get a feeling about possible changes due
to different trap geometries we will now calculate the collective frequencies
also for anharmonic one dimensional trap. The potential is given by
\beq
\label{1D_potential}
	V(x)=\frac{m\Omega^2}{2}x^2-\epsilon x^4.
\enq
From the TF limit we will get an equation relating chemical potential and
condensate boundary 
\beq
	\mu=\frac{m\Omega^2}{2}R^2-\epsilon R^4.
\enq
Using this result and continuing in the same manner as in previous 
subsection we
get an 1D analogue of Eq. (\ref{P_eq})
\beq
\label{1D_Peq}
	\frac{\partial}{\partial x}\left\{\left[(1-x^2)-\beta(1-x^4)
\right]\frac{\partial\delta n}{\partial x}\right\}+2\omega_N^2\delta n=0\, , 
\enq
where $\beta=\epsilon R^2/m\omega^2$. If the anharmonicity results 
from a Gaussian potential, then 
$\beta=R^2/8\sigma^2$. When $\beta=0$ the solutions have the form 
$\delta n=x^p\sum_{k=0}^N a_k x^k$, where $p=0$ or $1$.
If $p=0$ solutions are even and excitation frequencies are given by
$\omega_N^2=1/2\cdot N(N+1)$. If $p=1$ solutions are odd and corresponding
frequencies are given by $\omega_N^2=1/2\cdot (N+1)(N+2)$. For both even
and odd solutions $N$ must be an even integer.

	For the lowest excitations we calculate the shifted frequencies
$\omega^2(N=0,p=0)=0$, $\omega^2(N=0,p=1)=1-4\beta$,
$\,\omega^2(N=2,p=0)=3-\frac{30}{7}\beta$ and 
$\omega^2(N=2,p=1)=6-\frac{352}{45}\beta$. Especially we see that the 
corrections are always proportional to the dimensionless parameter $\beta$,
in the same way as in the spherically symmetric case. It is to be expected
that the calculated shifts will give correct order of magnitude estimates
even for more complicated trap geometries.

\section{Conclusions}\label{Conclusions}
In this paper we have demonstrated several ways by which the shallow
anharmonic potential changes the condensate properties. We have limited
our discussion mainly to Gaussian potentials, but most of the 
results can also be applied to traps having a quadratic anharmonic term.
Naturally, as an order of magnitude estimates, these result should be 
applicable also for other types of shallow traps.
It seems that in optical dipole traps the errors introduced by approximating 
the
potential as harmonic can be quite large. Harmonic model can predict
the condensate fraction poorly and with reasonable parameters the corrections
to the chemical potential and collective excitation frequencies can be
several percent. 
In an optical dipole trap the spin degree of freedom is not
necessarily frozen and the condensate should be described with a 
multicomponent spinor wavefunction~\cite{Miesner98,Stenger98,Ho98}.
Nevertheless, the results in this paper might prove to be useful also in 
studies of spinor condensates.

\section{Acknowledgments}
	The author acknowledges the Academy of Finland (project 43336)
and the National Graduate School on Modern Optics and Photonics for financial 
support. Discussions with Prof. K. A. Suominen are also greatly appreciated.

\begin{references}
\bibitem{Gaussian_comment}  In an optical dipole trap the 
potential is  proportional to intensity, and the  intensity of the focused 
laser beam is given by 
\begin{equation}
	I(r,z)\propto\frac{1}{1+(z/z_R)^2}\exp\left(-\frac{
	2r^2}{w_0^2\left(1+(z/z_R)^2\right)}\right).
\end{equation}
(See, for example, the review article, 
R. Grimm, M. Weidem\"{u}ller, and Y. Ovchinnikov, e-print physics/9902072,
and references therein.)
In this equation $w_0$ is the $1/e^2$ beam waist radius and  
$z_R$ is the Rayleigh range. While the intensity distribution is not Gaussian
in the $z$ direction, we expect that the results presented in this paper will 
give a rough estimate for the condensate properties in an optical dipole trap.
\bibitem{Chempot_comment} We have set the chemical potential to
be zero, implying zero energy for the ground state of the trap. This 
approximation  is valid only as long as the trap depth
is large compared to the ground state energy.
\end{references}
\end{document}